\documentclass[9pt,shortpaper,twoside,web]{ieeecolor}
\usepackage{generic}
\usepackage{cite}
\usepackage{amsmath,amssymb,amsfonts}
\usepackage{algorithmic}
\usepackage{graphicx}
\usepackage{textcomp}
\usepackage{multirow}
\def\BibTeX{{\rm B\kern-.05em{\sc i\kern-.025em b}\kern-.08em
    T\kern-.1667em\lower.7ex\hbox{E}\kern-.125emX}}
\begin{document}
\title{DeepMMSA: A Novel Multimodal Deep Learning Method for Non-small Cell Lung Cancer Survival Analysis}
\author{Yujiao Wu, Jie Ma, Xiaoshui Huang, Sai Ho Ling, \IEEEmembership{Member, IEEE}, and Steven Weidong Su, \IEEEmembership{Member, IEEE}
\thanks{Y.W., S.H.S., S.W.S., are now with Centre for Health Technologies, School of Biomedical Engineering, Faculty of Engineering and Information Technology, University of Technology Sydney, Sydney, NSW 2007, Australia. (e-mail: \{Yujiao.Wu, Steve.Ling, Steven.Su\}@uts.edu.au ).}
\thanks{X.H. is with ACRF Image X Institute, Faculty of Medicine and Health, University of Sydney, 1 Central Ave Australian Technology Park, Eveleigh NSW 2015, Australia. (e-mail: xiaoshui.huang@sydney.edu.au).}
\thanks{J.M. is with the Australian Artificial Intelligence Institute (AAII), Faculty of Engineering and IT, University of Technology Sydney, 15 Broadway, Ultimo NSW 2007, Australia. (e-mail: jie.ma-5@student.uts.edu.au).}}

\maketitle

\begin{abstract}
Lung cancer is the leading cause of cancer death worldwide. The critical reason for the deaths is delayed diagnosis and poor prognosis. With the accelerated development of deep learning techniques, it has been successfully applied extensively in many real-world applications, including health sectors such as medical image interpretation and disease diagnosis. By combining more modalities that being engaged in the processing of information, multimodal learning can extract better features and improve predictive ability. The conventional methods for lung cancer survival analysis normally utilize clinical data and only provide a statistical probability. To improve the survival prediction accuracy and help prognostic decision-making in clinical practice for medical experts, we for the first time propose a multimodal deep learning method for non-small cell lung cancer (NSCLC) survival analysis, named DeepMMSA. This method leverages CT images in combination with clinical data, enabling the abundant information hold within medical images to be associate with lung cancer survival information. We validate our method on the data of 422 NSCLC patients from The Cancer Imaging Archive (TCIA). Experimental results support our hypothesis that there is an underlying relationship between prognostic information and radiomic images. Besides, quantitative results showing that the established multimodal model can be applied to traditional method and has the potential to break bottleneck of existing methods and increase the the percentage of concordant pairs(right predicted pairs) in overall population by 4\%.

\end{abstract}


\section{Introduction}
\label{sec:introduction}
Lung cancer is the leading cancer killer in both men and women in the world, representing 19.4\%–27\% of all deaths from cancer~\cite{siegel2016cancer}. It can be broadly classified into non-small cell lung cancer (NSCLC), counting for 85\% and small cell lung cancer (SCLC), counting for the remaining 15\%. Lung cancer has a poor prognosis. The lung cancer five-year survival rate is lower than many other leading cancer sites, such as colorectal (64.5\%), breast (89.6\%), and prostate (98.2\%)~\cite{feuer2015national}. More than half of people with lung cancer die within one year of being diagnosed~\cite{feuer2015national}. Accurate assessment of the disease stage and survival time of lung cancer is essential in deciding the optimal plan and timing treatment for the clinicians.

Nowadays, a large fraction of interpretation of medical information is performed by medical experts. In terms of image interpretation by human experts, a lot of diagnostic errors appear in radiology. Approximately 20 million radiology reports contain clinically significant errors each year~\cite{brady2017error}. This limitation is mainly due to the subjectivity, the fatigue of the expert, the complexity of the image, as well as the extensive variations across different interpreters ~\cite{razzak2018deep}. Moreover, 2/3 of the world population lacks adequate access to radiology specialists; this would translate to 4.7 billion people.

Artificial intelligence (AI) is a promising tool that has shown its efficacy for diagnostic purposes\cite{liu2019comparison,zhu2018deeplung,xing2017deep}. With the rapid development in deep learning-based computer vision, its ability to recognize images or diagnose pictures even exceeds human ability \cite{ILSVRC15,imagenet_cvpr09,ILSVRCanalysis_ICCV2013}. The CNN is the most frequently used deep learning technique. In the last few years, image analysis by deep learning or CNNs has been utilized in low-dose lung CT for early diagnosis, which dramatically decreases the lung cancer mortality rate. To be specific, it enabled computer vision models to assist the doctors to detect suspicious pulmonary nodules or identify the location of the nodule, evaluate whole-lung/pulmonary malignancy, classify candidate nodules into benign or malignant, and predict the risk of lung cancer \cite{liao2019evaluate,gruetzemacher20183d,trajanovski2018towards, ardila2019end,zhu2018deeplung,ding2017accurate,li2020deepseed, riquelme2020deep}; in some cases, the models have reached competitive performance to doctors, and the accuracy even exceeds the doctors\cite{shin2012stacked,esteva2017dermatologist,gulshan2016development,litjens2017survey,xing2017deep}.

Although AI in combination with CT scans is a promising tool that has shown its utility for diagnostic purposes, it is rarely been used in predicting death, and possibly even determining death, which is a unique and challenging area that could be fraught with the same biases that affect analog physician-patient interactions. Current research for NSCLC survival analysis is largely based on the statistical analysis of clinical data. Traditional approaches generally utilize clinical information such as age, clinical TNM stage, gender information, etc. In the work of\cite{wang2019deep}, which uses CT images in survival analysis. But the model only learns 2D features from each tumor image slice separately, afterward, the averaged features from all image slices are calculated for every single patient and overlook the third-dimensional properties of the tumor. Moreover, the overall performance of currently available works for survival analysis is considerably low. 

By contrast, deep learning has the potential to reduce diagnostic errors and overcome human limitations. It is worth developing a fully automated deep learning based method for NSCLC survival prediction based on CT images and clinical data to explore whether the deep learning techniques have the ability to extract useful information from CT images and clinical data to predict death time. In our work, considering the 3D nature of CT images, to reveal the underlying relation between prognostic information and CT images, fully utilize the potential of the prognostic power existing in the radiomic data, we design a 3D multimodal deep learning based model for NSCLC survival analysis. Since combining CT image and clinical data could provide comprehensive and supplementary information to describe the cancer status, this method has the potential to effectively increase prediction accuracy. To our best knowledge, this is the first time to develop a multimodal deep learning method which fuse the 3D features extracted from lung CT images with the features extracted from clinical data for death prediction. 

After we designed the model, we use the concordance statistic to evaluate our method. Concordance, or synonymously the C-index, is one of the most popular performance measures in accessing survival models~\cite{therneau20201,kremers2007concordance,steck2008ranking} . It measures the discrimination ability of a model in survival analysis\cite{heller2016estimating}. C-index refers to the proportion of pairs whose predicted results are consistent with actual results among all patient pairs. For instance, the C-index is increased by 1\% implies that if a survival 
analysis is made for a population of one million people, the correct prediction will increase the number of people by four thousand. Quantitative results on the NSCLC- Radiomics data show that the proposed method can make a contribution to the state-of-the-art methods and can increase C-index by 4\%. Revealing that discrimination of proposed model is improved. And can provide a more accurate prognostic decision-making in future clinical practice. The results of ablation experiments show that using multiple modalities improves the biggest marginal performance compared with using a single modality.

To conclude, we mainly have the following contributions: 
\begin{itemize}
	\item The proposed method overcomes the weakness of traditional non-parametric methods (KM etc.), which cannot incorporate multiple variables. 
	\item Experiment results show that our method could improve the performance of SOTA methods (Cox-Time etc.) by 4\% on C-index. 
	\item The first attempt to reconstruct a deep 3D convolutional-based model and using images for survival analysis. 
	\item Experiment results support our hypothesis and reveal the underlying relation between prognostic information and radiomic images. 
	\item Our multimodal method can provide more accurate survival analysis with sufficient granularity for personalized prognosis and decision-making in combination with SOTA methods. 
\end{itemize}

\section{Preliminary Knowledge}
\label{sec:Preliminary Knowledge}
In this section, we provide a brief survey on multimodal deep learning, 3D ResNets, survival analysis, and survival analysis methods developed in recent years.
\subsection{Notations}
For notational clarity, we hereby define the key symbols and their meanings in Table \ref{notation}.

\begin{table}[h]
	\caption{Notations.}
	\label{notation}
	\begin{center}
	\begin{tabular}{c|c}
		\hline
		Symbol  & Definition \\
		\hline
		$r_i$       & $i$-th 3D radiology image      \\
		$c_i$       & $i$-th clinical information  \\
		$y_i$      & Time $i$-th event of interest or censoring occurred      \\\
		$\hat{y_i}$   & predicted or estimated $y_i$    \\
		$e_i$     &  $i$-th event, 1 for uncensored, 0 for censored    \\
		$I(x)$    & 1 if $x=True$ else 0 \\
		\hline
	\end{tabular}
    \end{center}
\end{table}

\subsection{Multimodal Deep Learning}
Multimodal deep learning is the novel framework using deep learning to learn from multiple modalities, such as text, images, and audio ~\cite{ngiam2011multimodal}. In medical applications, multiple types of data are related to each patient, including clinical information, radiology images, physician note, medication, to name a few. Thus, when data comes from different sources, the approach of multimodal deep learning can help to understand and extract more useful information.

\subsection{3D-ResNet}
The residual neural network (ResNet)\cite{he2016deep} can handle gradient vanishing or exploding problems in deeper neural networks training, especially in computer vision. The core concept of ResNet is to construct a basic network block in which the output adds up with input. To handle 3D image input, we can simply increase kernel shape from 2 dimensions to 3 dimensions (height, length, and depth) in convolution layers.

\subsection{Survival Analysis}
Survival analysis is a widely used technique to analyze time-to-event data (e.g., student dropout, cancer survival, admission to hospital, disease recurrence, etc.)\cite{lee2003statistical}. Traditional statistical methods for survival analysis normally contains three options for modeling the survival function: non-parametric methods with no distribution of survival curve predefined (e.g., Kaplan-Meier\cite{kaplan1958nonparametric}, Nelson-Aalen\cite{nelson1972theory,aalen1978nonparametric}), semi-parametric methods such as the Cox proportional hazards model\cite{cox1972regression} (Cox regression) which is most commonly applied, and parametric methods with distribution predefined (e.g., Linear regression, Weibull distribution). Besides, machine learning methods such as survival trees, neural network, Cox-time\cite{kvamme2019time}, DeepHit\cite{lee2018deephit}, CoxCC\cite{kvamme2019time}, PC-Hazard\cite{kvamme2019continuous} and ensemble methods, to name just a few, are also applied in survival analysis.

Due to the existence of censored survival data (usually right censored), the standard evaluation indexes for regression, such as mean square error (MSE) and $R^2$, do not fit for quantifying the performance of survival analysis. The most important evaluation index is C-index, which can evaluate uncensored instances and censored instances together. 
\begin{equation}
\text{C-index}=\frac{\sum_{i,j}I(\hat{y_i}<\hat{y_j}|e_i=1,y_i<y_j)}{\sum_{i,j}I(y_i<y_j|e_i=1)}
\end{equation}
Besides, as a regression problem, we also use the mean absolute error (MAE) over uncensored instances to evaluate our experiments.
\begin{equation}
\text{MAE}=\frac{1}{\sum_i I(e_i=1)}\sum_{i=1}^N(e_i|y_i-\hat{y_i}|)
\end{equation}

\subsection{Related Work}
Conventional survival analysis for NSCLC is a set of modeling procedures that only harness clinical data, and it measures time to an event. For example, in the specific area of NSCLC survival analysis, \cite{janssen1998variation} studied the differences in the prognosis for European adult patients with lung cancer, by age, country, etc. from 1985 to 1989 with simple statistical methods, such as life-table. The work of\cite{port2003tumor} shows that tumor size within stage IA is an important predictor of survival with the Kaplan-Meier and Cox regression model. \cite{gyorffy2013online} developed an online survival analysis tool that capable of uni- and multivariate analyzing (e.g., Kaplan-Meier survival plot, Cox regression analysis) with 1,715 samples of ten independent datasets. Recently, some work has been proposed to use deep learning in survival analysis. DeepConvSurv\cite{zhu2016deep}, which for the first time developed a 2D deep CNN for survival analysis with pathological images. The study of \cite{chaddad2017predicting} uses the random forest model to analyze the manually extracted features from radiology images combined with age to predict a binary classification of survival time. \cite{wang2019deep} used CNN to extract and then average 2D features from CT image slices in survival analysis, which may lose important information such as the spatial information that exists in tumors. \cite{cui2020deep} used a deep neural network to learn cellular features from biomarkers and Cox proportional hazards model to do survival analysis. DeepLung\cite{zhu2018deeplung} use 3D CT images and 3D CNN for nodule detection and classification, the success of which inspired us to extract 3D features from CT images for survival analysis.

In summary, the early-stage research for NSCLC survival analysis tends to find specific features to predict the survival curve, most of which using the Kaplan-Meier and Cox regression model. Then with the rising of deep learning in recent years, deep models tend to be used for analyzing more features and more types of data, such as image and text, in NSCLC survival analysis.

In this paper, for individualized NSCLC survival prediction, we propose a multimodal deep learning method including 3D-ResNet, which could fully utilize and analyze the information gathered from all types of data sources, such as CT images and clinical information.

\section{Methodology}
In this section, we describe details about the proposed DeepMMSA. As far as we know, this is the first work to use a multimodal deep learning method to process CT images together with clinical information for NSCLC survival analysis.

\subsection{Structure of DeepMMSA}
\begin{figure*}[thpb] 
	\begin{center}
		\includegraphics[width=1.0\linewidth]{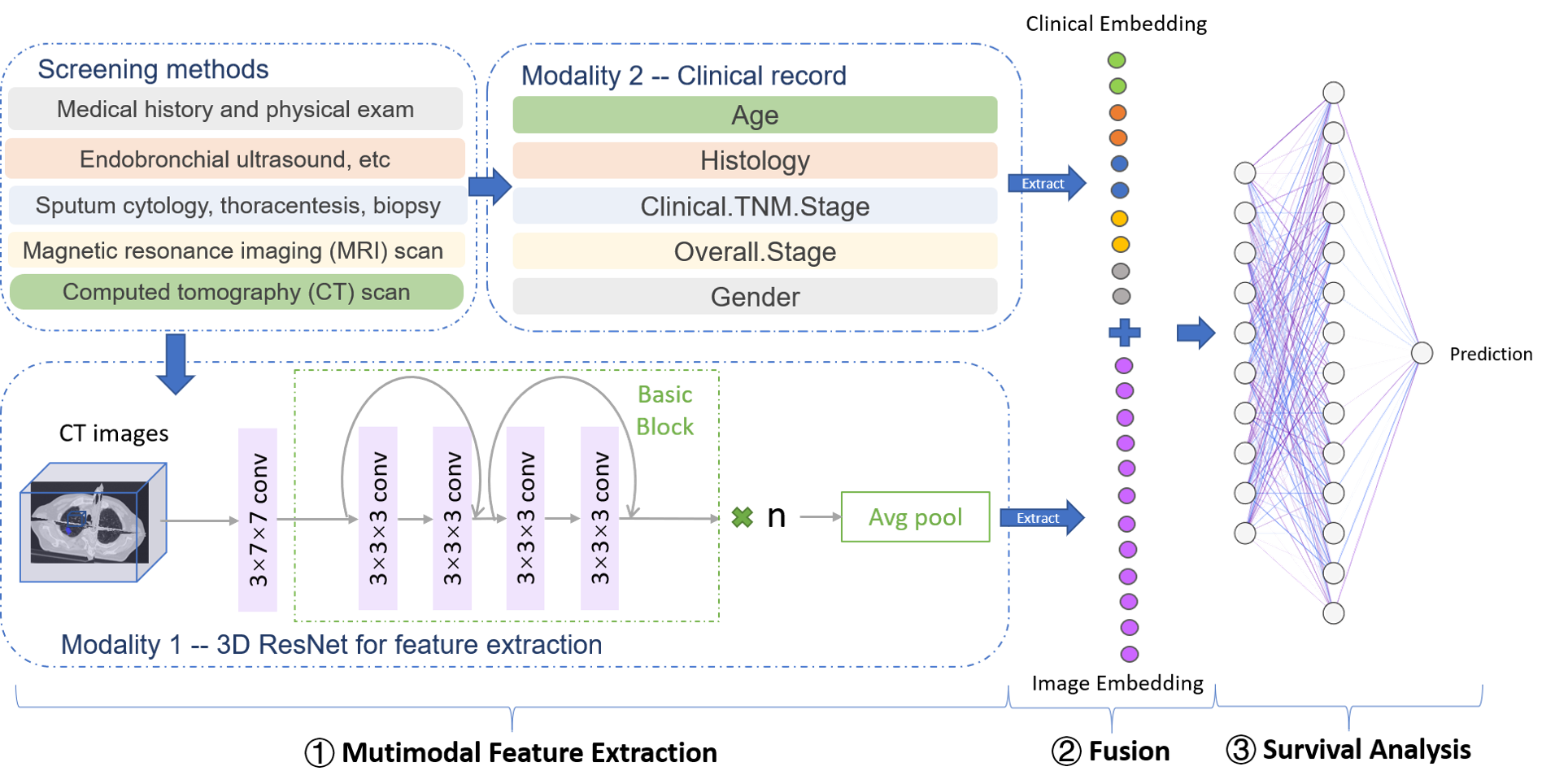}
	\end{center}
	\caption{The workflow of deepMMSA. DeepMMSA mainly has three module: (1)First, it employs the 3D-ResNet in combination with plain networks for multimodal feature extraction; (2) Then, it uses simple feature fusion method (early fusion) for multimodal fusion; (3) Lastly, during desicion making stage, plain neural network is designed for the survival prediction. }
	\label{RES}
\end{figure*}

Inspired by the recent successful applications of CNNs in NSCLC diagnosis and other image recognition tasks, we proposed a DeepMMSA that harnesses the strategies of analyzing the combined information from multiple modalities. As shown in Fig. \ref{RES}, we use the tumor region of interests (ROIs), which is the lesions on CT scans as low-level image input. Motivated by the work of\cite{zhu2018deeplung}, 3D CNNs were used to extract features from all three-dimensional directions within the tumor volume. Meanwhile, we integrate the clinical information through a 27-dimension high-level clinical layer, the input of which includes screening test results such as clinical TNM stage, overall stage, histology, gender, age of the patient. The clinical layer is embedded in the hidden layer directly.

DeepMMSA method consists of three modules: 
{
	\begin{itemize}
		\item Multimodal feature (CT image features and Clinical record features) extraction.
		\item Multimodal feature fusion.
		\item Survival analysis.
	\end{itemize}
}

\subsubsection{Multimodal Feature Extraction}

\paragraph{CT Images Feature Extraction with 3D-ResNet}

\begin{figure}[!htb]
	\centering
	\includegraphics[width=0.95\columnwidth]{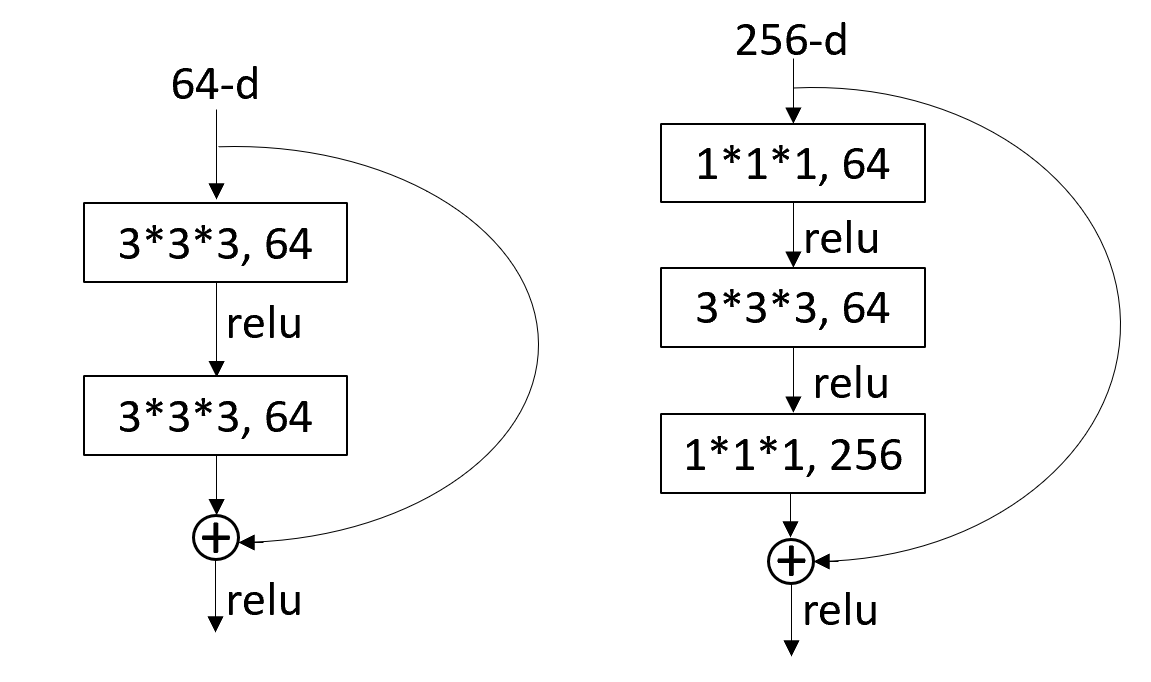}
	\caption{The deep reseidual function $\mathbf{F}$ of 3D-ResNet. The left figure is a building block for 3D-ResNet with layers of 18 and 34. The right figure is a bottleneck building block for 3D-ResNet with layers of 50, 101, and 152.}
	\label{resblock}
\end{figure}

As is shown in Fig.\ref{RES} our method requires two multimodal inputs, CT images, and clinical data from the according patient. To be specific, CT images are introduced to the radiomic embedding layer and clinical data are introduced to the clinical embedding layer. We propose 3D-ResNets as our network structure for the low-level image feature learning. As shown in Fig. \ref{RES} and Fig. \ref{resblock}, 3D-ResNets can be built by basic blocks or 'bottleneck' building blocks, the number of which may vary from 18 to 152 in the whole network. For instance, 3D-ResNet-18 contains four basic blocks, and each block contains four convolutional layers (conv1-conv4). Compared with 2D ResNet, features extracted in all three-dimensional directions within the tumor volume with 3D ResNet, therefore taking the third-dimensional spatial information into consideration\cite{aerts2014decoding}. It worth noting that, in order to eliminate the vanishing and exploding gradients types of problems in the very deep neural network, we add extra shortcut connections in our model. The advantage of such residual learning is that it enables the reuse of features. Since our dataset is relatively small compared to other general datasets for image recognition, we use data argumentation techniques before importing the data into the model. 

The formation of the residual building block is,
\begin{equation}
o=\mathbf{F}(x)+x
\end{equation}
in which $\mathbf{F}$ is the deep residual function, $x,o$ are the input and output.

\paragraph{Clinical Record Feature Extraction}

The clinical embedding layers introduce clinical data to the network separately aims to capture the survival information indicated within clinical data. For clinical data, a neural network with two hidden layers was proposed to extract features. As is shown in Fig.\ref{RES}, the 27-dimension non-image features were extracted by a network, which in combination with the image features is processed in the later fusion stage.

\subsubsection{Multimodal Feature Fusion}
During the feature fusion stage, multimodal features from CT images and clinical records are difficult to be directly fused together. This is due to the features from different modalities have different scales or statistical properties. Thus, to solve this problem, we applied the Batch Normalization (BN) technique to adjust the mean and variance of extracted features in each modality before the fusion procedure. Given the features $z_1,z_2,...,z_m$ over a batch, $\hat{z_i}$ is calculated as:
\begin{equation}
\hat{z_i}=\gamma_i\frac{z_i-\mu_i}{\sqrt{\sigma_i^2+\epsilon}}+\beta_i
\end{equation}
where $\gamma_i,\beta_i$ are the parameters to be learned, $\mu_i$ is the mean value of $z_i$ over the batch, $\sigma_i$ is the standard deviation of $z_i$ over the batch, $\epsilon$ is set to a very small number such as $10^{-8}$.

\subsubsection{Survival Analysis}
Survival analysis is the last module in the method and is intended as an overall feature analysis to get the final survival time prediction. In this module, to meet the requirements for a specific problem, any survival analysis model can be used. It can be models either from traditional statistical methods, such as the Kaplan-Meier model, Cox regression model, or from novel machine learning methods, such as survival trees, and deep neural networks. All the mentioned models can be used to analyze the input from multimodal features in our proposed fusion layer. In this work, we define the survival time as the label, and use the one hidden layer neural network with one dimension output layer for the regression setting and overall optimization convenience. Additionally, in order to achieve the same scale between ground truth and the output in our model, the Sigmoid function is used as a normalization technique to process the output.

\begin{equation}
\hat{y_i}=\frac{1}{1+e^{-f(x_i)}}
\end{equation}
and we use MSE loss function and $L2$ regularization penalty term as the objective function which is defined as:
\begin{equation}
\text{minimize\ } L=\frac{1}{N}\sum_{i=1}^N (y_i-\hat{y_i})^2 + \lambda \sum_{j=1}^M w_j^2
\end{equation}
where $w_j$ is the model parameter and the total number is M. $\lambda$ is the penalty coefficient.

\section{Experiments}
We conduct extensive experiments based on NSCLC patients from TCIA to validate the performance of our proposed method DeepMMSA with several state-of-the-art methods in terms of the prediction accuracy for the survival time for each patient. Besides, we also evaluate the prediction result by C-index. Afterward, we perform several ablation experiments regarding different network structures to determine the best structure. 

\subsection{Dataset}

In this work, we considered 422 NSCLC patients from TCIA to assess the proposed method. For these patients pretreatment CT scans, manual delineation by a radiation oncologist of the 3D volume of the gross tumor volume and clinical outcome data are available\cite{clark2013cancer}. The corresponding clinical data are also available in the same collection. The patients who had neither survival time nor event status were excluded from this work. 

\subsection{Data Preprocessing}
For CT images, we resize the raw data which is the 3D volume of the primary gross tumor volume into $96*96*8$. After that, we transform the range linearity into [0,1]. Then, to prevent overfitting, we perform data argumentation which includes three methods: rotate, swap, and flip. Then we get $422*8=3376$ samples, among which there are $373*8=2984$ uncensored samples and $49*8=392$ censored samples. 

Clinical data contains categorical data and non-categorical data. Firstly, missing value is a common problem in medical data and may pose difficulties for data analyzing and modeling. Specifically, in our dataset, the 'age' category contains a few missing values. After observing the data, we find that the age of patients in the dataset is close to each other. Thus, we impute the mean value and fill it into the missing value. Afterward, in order to fit into our model, we use the one-hot encoder to encode categorical data into numbers, which allows the representation of categorical data to be more expressive. 

Then, we use the min-max feature scaling method and standard score method to perform data normalization, such as age and survival time. For input $x$, the min-max feature scaling method's output is:
\begin{equation}
x'=\frac{x-x_{min}}{x_{max}-x_{min}}
\end{equation}
and the standard score method's output is:
\begin{equation}
x'=\frac{x-mean(x)}{std(x)}
\end{equation}
where std is the standard deviation.

For a single patient with multiple tumors, we select the primary gross tumor volume ('GTV-1') to be processed in our work, while other tumors such as secondary tumor volumes denoted as 'GTV-2', 'GTV-3' to name just a few, which were occasionally present, were not considered in our work.

\subsection{Experiment Setup}
We train and evaluate the method on the NSCLC-Radiomic dataset following 5-fold cross-validation with the patient-level split. We divided the dataset into training, validation, and testing data into 6:2:2 respectively. For hyperparameters tuning such as the penalty coefficient, we use the validation dataset to fine-tune and get the optimized hyperparameters. In the training process, we use 200 epochs in total with Adam as the optimizer. The batch size parameter is set as 64. The initial learning rate is set as 0.001, and then the learning rate is decayed by 0.5 in every 40 epochs.

Since we use survival time as the label, not cumulative hazard. In the training and validation process, we only use the uncensored data for precise survival time and objective function calculation, and in the testing process, we use all data for C-index evaluation and uncensored data for MAE evaluation. 

Since this is the first work to use a multimodal method for NSCLC survival analysis, we implement several state-of-the-art survival analysis methods as baselines to compare with our work. The baseline methods include Cox-time\cite{kvamme2019time}), DeepHit\cite{lee2018deephit}, CoxCC\cite{kvamme2019time}, PC-Hazard\cite{kvamme2019continuous} and the regular cox regression.

\subsection{Ablation Study}
To find an optimal network for our problem, we consider performing ablation experiments based on the following four aspects of network architecture: 
{
	\begin{itemize}
		\item How the depth of Resnet affect the performance? Which 3D structure is the best?
		\item Whether multiple modalities outperform single modality?
		\item What is the best ratio set between image data and clinical data in the fusion stage?
		\item Whether hidden layers should be added in the survival analysis stage?
	\end{itemize}
}

Firstly, the evaluation of different depths of Resnet and whether multiple modalities are better than a single modality are conducted. By using one hidden layer in the survival analysis network and fixing the ratio of multiples modalities features in the fusion stage to 512:27, four different structures were tested. Table \ref{structures} shows that our model of r3d34 structure with multiple modalities achieves the best performance. All experimental results support our assumption and show the effectiveness of using multiple modalities in survival analysis. 

Moreover, using the best structure r3d34, we step further to observe the effects of changing the ratio between modalities with and without the hidden layer. It can be done by changing the number of perceptrons for image and non-image features. The comparison results list in Table \ref{ratio} shows that setting the ratio between images and non-image modalities as 25:27 achieves the best performance. Besides, adding hidden layers in the survival analysis module can further improve the performance. After ablation experiments, the best architecture of using r3d34 structure, setting the ratio between multiple modalities as 25:27 with one hidden layer outperform the other structures in this work. 

The results indicate that NSCLC prognosis information can be learned from CT images. Using multimodal structure, more information can be jointly extracted and learned from various information sources and non-linearly transformed into a deep network, therefore improve the overall prediction accuracy. Furthermore, using multiple modalities improves the biggest marginal performance compared with using single modalities.

\begin{table}
	\caption{To evaluate the performance of different ResNet structure and effect of whether using multiple modalities.}
	\label{structures}
	\centering
	\begin{tabular}{l|ll|ll}
		\hline
		\multirow{2}{*}{} & \multicolumn{2}{l|}{CT images} & \multicolumn{2}{l}{Multi-modality} \\ \cline{2-5} 
		& Loss           & C-index          & Loss             & C-index            \\ \hline
		r3d18                  & 0.1023           & 0.5782           & 0.0847             & 0.6287            \\
		r3d34                  & 0.0975           & 0.5942           & \textbf{0.0757}             & \textbf{0.6490}            \\
		r3d50                  & 0.1026           & 0.5804           & \text{0.0760}              & 0.6375            \\
		r3d101                 & 0.1071           & 0.5660           & 0.0795             & 0.6142            \\ \hline
	\end{tabular}
\end{table}

\begin{table}
	\caption{To evaluate the effects of different ratio between modalities features in fusion procedure and the performance of survival analysis neural network with or without hidden layer.}
	\label{ratio}
	\centering
	\begin{tabular}{l|ll|ll}
		\hline
		\multirow{2}{*}{} & \multicolumn{2}{l|}{Hidden} & \multicolumn{2}{l}{No hidden} \\ \cline{2-5} 
		& Loss          & C-index         & Loss           & C-index          \\ \hline
		512:27            & 0.0757             & 0.6490           & 0.0760              & 0.6376           \\
		100:27            & 0.0745             & 0.6512           & 0.0755              & 0.6421            \\
		25:27             & \textbf{0.0739}            & \textbf{0.6580}          & 0.0761             & 0.6450            \\
		5:27              & 0.0765             & 0.6403           & 0.0793              & 0.6215            \\ \hline
	\end{tabular}
\end{table}


\subsection{Results}
\begin{figure}[!htb]
	\centering
	\includegraphics[width=1.0\columnwidth]{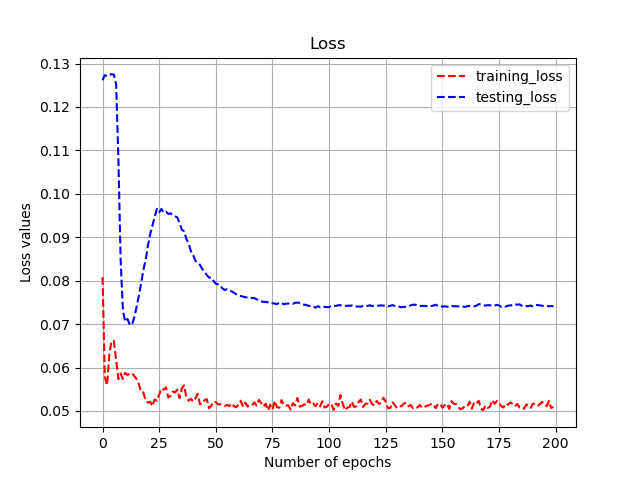}
	\caption{Training and testing process.}
	\label{loss}
\end{figure}

The loss in the training and testing process is shown in Fig. \ref{loss}.

For a fair comparison, we perform five state-of-the-art methods as baseline methods from previous work for survival analysis. As shown in Table \ref{baseline}, we compare our method and with the other five baseline models. The results show that our method achieves the best performance. Besides, the results verify that compared to the methods(Cox-time, DeepHit, etc.) which simply use clinical data, DeepMMSA shows its superior of effectively extracting the supplementary information from multiple modalities and can significantly improve the prediction accuracy.




\begin{table}
	\caption{Results vs baselines.}
	\label{baseline}
	\centering
	\begin{tabular}{l|ll}
		\hline
		Model          & MAE & C-index \\ \hline
		Cox-time          & 0.183   & 0.6152       \\
		Cox regression       & 0.204   & 0.6009   \\
		CoxCC       & 0.183   & 0.6120      \\
		PC-Hazard  & 0.191   & 0.6094     \\
		DeepHit   & 0.183   & 0.6133       \\
		DeepMMSA      & \textbf{0.162}   & \textbf{0.6580}       \\ \hline
	\end{tabular}
\end{table}


\section{Conclusion and Future Work} 
In this paper, we proposed a fully automated end-to-end multimodal deep network method for NSCLC survival analysis. Our method can learn complementary representations from the CT image and non-image clinical data modalities. Extensive experimental results show that DeepMMSA outperforms conventional methods that use a single source of information alone. But there is still some future work to do. There are some potential ways to improve the performance of the proposed method. 
Since there are three basic modules for multimodal deep learning survival analysis framework, we consider to made improvements based on the following three aspects:  
\begin{itemize}
	\item Provide more complementary information by adding more modalities to improve the performance, such as e-nose based diagnosis\cite{nakhleh2017diagnosis} time series data, etc., and try to fully exploit the inherent correlations across multiple modalities.
	\item Perform different multimodal fusion approaches, such as decision fusion and hybrid fusion method, etc.
	\item In the survival analysis module, theoretically, we can use any survival analysis model, such as cox-time, deepsurv, to improve the performance of our proposed framework.
\end{itemize}

{\small
\bibliographystyle{ieeetr}
\bibliography{main}

\begin{thebibliography}{10}

\bibitem{siegel2016cancer}
R.~L. Siegel, K.~D. Miller, and A.~Jemal, ``Cancer statistics, 2016,'' {\em CA:
  a cancer journal for clinicians}, vol.~66, no.~1, pp.~7--30, 2016.

\bibitem{feuer2015national}
E.~Feuer, N.~Howlader, A.~Noone, M.~Krapcho, J.~Garshell, D.~Miller, {\em
  et~al.}, ``National cancer institute seer cancer statistics review,'' {\em
  National Cancer Institute}, vol.~103, no.~7, pp.~1975--2012, 2015.

\bibitem{brady2017error}
A.~P. Brady, ``Error and discrepancy in radiology: inevitable or avoidable?,''
  {\em Insights into imaging}, vol.~8, no.~1, pp.~171--182, 2017.

\bibitem{razzak2018deep}
M.~I. Razzak, S.~Naz, and A.~Zaib, ``Deep learning for medical image
  processing: Overview, challenges and the future,'' in {\em Classification in
  BioApps}, pp.~323--350, Springer, 2018.

\bibitem{liu2019comparison}
X.~Liu, L.~Faes, A.~U. Kale, S.~K. Wagner, D.~J. Fu, A.~Bruynseels,
  T.~Mahendiran, G.~Moraes, M.~Shamdas, C.~Kern, {\em et~al.}, ``A comparison
  of deep learning performance against health-care professionals in detecting
  diseases from medical imaging: a systematic review and meta-analysis,'' {\em
  The lancet digital health}, vol.~1, no.~6, pp.~e271--e297, 2019.

\bibitem{zhu2018deeplung}
W.~Zhu, C.~Liu, W.~Fan, and X.~Xie, ``Deeplung: Deep 3d dual path nets for
  automated pulmonary nodule detection and classification,'' in {\em 2018 IEEE
  Winter Conference on Applications of Computer Vision (WACV)}, pp.~673--681,
  IEEE, 2018.

\bibitem{xing2017deep}
F.~Xing, Y.~Xie, H.~Su, F.~Liu, and L.~Yang, ``Deep learning in microscopy
  image analysis: A survey,'' {\em IEEE Transactions on Neural Networks and
  Learning Systems}, vol.~29, no.~10, pp.~4550--4568, 2017.

\bibitem{ILSVRC15}
O.~Russakovsky, J.~Deng, H.~Su, J.~Krause, S.~Satheesh, S.~Ma, Z.~Huang,
  A.~Karpathy, A.~Khosla, M.~Bernstein, A.~C. Berg, and L.~Fei-Fei, ``{ImageNet
  Large Scale Visual Recognition Challenge},'' {\em International Journal of
  Computer Vision (IJCV)}, vol.~115, no.~3, pp.~211--252, 2015.

\bibitem{imagenet_cvpr09}
J.~Deng, W.~Dong, R.~Socher, L.-J. Li, K.~Li, and L.~Fei-Fei, ``{ImageNet: A
  Large-Scale Hierarchical Image Database},'' in {\em CVPR09}, 2009.

\bibitem{ILSVRCanalysis_ICCV2013}
O.~Russakovsky, J.~Deng, Z.~Huang, A.~C. Berg, and L.~Fei-Fei, ``Detecting
  avocados to zucchinis: what have we done, and where are we going?,'' in {\em
  International Conference on Computer Vision (ICCV)}, 2013.

\bibitem{liao2019evaluate}
F.~Liao, M.~Liang, Z.~Li, X.~Hu, and S.~Song, ``Evaluate the malignancy of
  pulmonary nodules using the 3-d deep leaky noisy-or network,'' {\em IEEE
  transactions on neural networks and learning systems}, vol.~30, no.~11,
  pp.~3484--3495, 2019.

\bibitem{gruetzemacher20183d}
R.~Gruetzemacher, A.~Gupta, and D.~Paradice, ``3d deep learning for detecting
  pulmonary nodules in ct scans,'' {\em Journal of the American Medical
  Informatics Association}, vol.~25, no.~10, pp.~1301--1310, 2018.

\bibitem{trajanovski2018towards}
S.~Trajanovski, D.~Mavroeidis, C.~L. Swisher, B.~G. Gebre, B.~S. Veeling,
  R.~Wiemker, T.~Klinder, A.~Tahmasebi, S.~M. Regis, C.~Wald, {\em et~al.},
  ``Towards radiologist-level cancer risk assessment in ct lung screening using
  deep learning,'' {\em arXiv preprint arXiv:1804.01901}, 2018.

\bibitem{ardila2019end}
D.~Ardila, A.~P. Kiraly, S.~Bharadwaj, B.~Choi, J.~J. Reicher, L.~Peng, D.~Tse,
  M.~Etemadi, W.~Ye, G.~Corrado, {\em et~al.}, ``End-to-end lung cancer
  screening with three-dimensional deep learning on low-dose chest computed
  tomography,'' {\em Nature medicine}, vol.~25, no.~6, pp.~954--961, 2019.

\bibitem{ding2017accurate}
J.~Ding, A.~Li, Z.~Hu, and L.~Wang, ``Accurate pulmonary nodule detection in
  computed tomography images using deep convolutional neural networks,'' in
  {\em International Conference on Medical Image Computing and
  Computer-Assisted Intervention}, pp.~559--567, Springer, 2017.

\bibitem{li2020deepseed}
Y.~Li and Y.~Fan, ``Deepseed: 3d squeeze-and-excitation encoder-decoder
  convolutional neural networks for pulmonary nodule detection,'' in {\em 2020
  IEEE 17th International Symposium on Biomedical Imaging (ISBI)},
  pp.~1866--1869, IEEE, 2020.

\bibitem{riquelme2020deep}
D.~Riquelme and M.~A. Akhloufi, ``Deep learning for lung cancer nodules
  detection and classification in ct scans,'' {\em AI}, vol.~1, no.~1,
  pp.~28--67, 2020.

\bibitem{shin2012stacked}
H.-C. Shin, M.~R. Orton, D.~J. Collins, S.~J. Doran, and M.~O. Leach, ``Stacked
  autoencoders for unsupervised feature learning and multiple organ detection
  in a pilot study using 4d patient data,'' {\em IEEE transactions on pattern
  analysis and machine intelligence}, vol.~35, no.~8, pp.~1930--1943, 2012.

\bibitem{esteva2017dermatologist}
A.~Esteva, B.~Kuprel, R.~A. Novoa, J.~Ko, S.~M. Swetter, H.~M. Blau, and
  S.~Thrun, ``Dermatologist-level classification of skin cancer with deep
  neural networks,'' {\em nature}, vol.~542, no.~7639, pp.~115--118, 2017.

\bibitem{gulshan2016development}
V.~Gulshan, L.~Peng, M.~Coram, M.~C. Stumpe, D.~Wu, A.~Narayanaswamy,
  S.~Venugopalan, K.~Widner, T.~Madams, J.~Cuadros, {\em et~al.}, ``Development
  and validation of a deep learning algorithm for detection of diabetic
  retinopathy in retinal fundus photographs,'' {\em Jama}, vol.~316, no.~22,
  pp.~2402--2410, 2016.

\bibitem{litjens2017survey}
G.~Litjens, T.~Kooi, B.~E. Bejnordi, A.~A.~A. Setio, F.~Ciompi, M.~Ghafoorian,
  J.~A. Van Der~Laak, B.~Van~Ginneken, and C.~I. S{\'a}nchez, ``A survey on
  deep learning in medical image analysis,'' {\em Medical image analysis},
  vol.~42, pp.~60--88, 2017.

\bibitem{wang2019deep}
S.~Wang, Z.~Liu, Y.~Rong, B.~Zhou, Y.~Bai, W.~Wei, M.~Wang, Y.~Guo, and
  J.~Tian, ``Deep learning provides a new computed tomography-based prognostic
  biomarker for recurrence prediction in high-grade serous ovarian cancer,''
  {\em Radiotherapy and Oncology}, vol.~132, pp.~171--177, 2019.

\bibitem{therneau20201}
T.~Therneau and E.~Atkinson, ``1 the concordance statistic,'' 2020.

\bibitem{kremers2007concordance}
W.~K. Kremers, ``Concordance for survival time data: fixed and time-dependent
  covariates and possible ties in predictor and time,'' {\em Mayo Foundation},
  2007.

\bibitem{steck2008ranking}
H.~Steck, B.~Krishnapuram, C.~Dehing-Oberije, P.~Lambin, and V.~C. Raykar, ``On
  ranking in survival analysis: Bounds on the concordance index,'' in {\em
  Advances in neural information processing systems}, pp.~1209--1216, Citeseer,
  2008.

\bibitem{heller2016estimating}
G.~Heller and Q.~Mo, ``Estimating the concordance probability in a survival
  analysis with a discrete number of risk groups,'' {\em Lifetime data
  analysis}, vol.~22, no.~2, pp.~263--279, 2016.

\bibitem{ngiam2011multimodal}
J.~Ngiam, A.~Khosla, M.~Kim, J.~Nam, H.~Lee, and A.~Y. Ng, ``Multimodal deep
  learning,'' in {\em ICML}, 2011.

\bibitem{he2016deep}
K.~He, X.~Zhang, S.~Ren, and J.~Sun, ``Deep residual learning for image
  recognition,'' in {\em Proceedings of the IEEE conference on computer vision
  and pattern recognition}, pp.~770--778, 2016.

\bibitem{lee2003statistical}
E.~T. Lee and J.~Wang, {\em Statistical methods for survival data analysis},
  vol.~476.
\newblock John Wiley \& Sons, 2003.

\bibitem{kaplan1958nonparametric}
E.~L. Kaplan and P.~Meier, ``Nonparametric estimation from incomplete
  observations,'' {\em Journal of the American statistical association},
  vol.~53, no.~282, pp.~457--481, 1958.

\bibitem{nelson1972theory}
W.~Nelson, ``Theory and applications of hazard plotting for censored failure
  data,'' {\em Technometrics}, vol.~14, no.~4, pp.~945--966, 1972.

\bibitem{aalen1978nonparametric}
O.~Aalen, ``Nonparametric inference for a family of counting processes,'' {\em
  The Annals of Statistics}, pp.~701--726, 1978.

\bibitem{cox1972regression}
D.~R. Cox, ``Regression models and life-tables,'' {\em Journal of the Royal
  Statistical Society: Series B (Methodological)}, vol.~34, no.~2,
  pp.~187--202, 1972.

\bibitem{kvamme2019time}
H.~Kvamme, {\O}.~Borgan, and I.~Scheel, ``Time-to-event prediction with neural
  networks and cox regression,'' {\em Journal of machine learning research},
  vol.~20, no.~129, pp.~1--30, 2019.

\bibitem{lee2018deephit}
C.~Lee, W.~Zame, J.~Yoon, and M.~van~der Schaar, ``Deephit: A deep learning
  approach to survival analysis with competing risks,'' in {\em Proceedings of
  the AAAI Conference on Artificial Intelligence}, vol.~32, 2018.

\bibitem{kvamme2019continuous}
H.~Kvamme and {\O}.~Borgan, ``Continuous and discrete-time survival prediction
  with neural networks,'' {\em arXiv preprint arXiv:1910.06724}, 2019.

\bibitem{janssen1998variation}
M.~Janssen-Heijnen, G.~Gatta, D.~Forman, R.~Capocaccia, J.~Coebergh, E.~W.
  Group, {\em et~al.}, ``Variation in survival of patients with lung cancer in
  europe, 1985--1989,'' {\em European Journal of Cancer}, vol.~34, no.~14,
  pp.~2191--2196, 1998.

\bibitem{port2003tumor}
J.~L. Port, M.~S. Kent, R.~J. Korst, D.~Libby, M.~Pasmantier, and N.~K.
  Altorki, ``Tumor size predicts survival within stage ia non-small cell lung
  cancer,'' {\em Chest}, vol.~124, no.~5, pp.~1828--1833, 2003.

\bibitem{gyorffy2013online}
B.~Gyorffy, P.~Surowiak, J.~Budczies, and A.~Lanczky, ``Online survival
  analysis software to assess the prognostic value of biomarkers using
  transcriptomic data in non-small-cell lung cancer,'' {\em PloS one}, vol.~8,
  no.~12, pp.~e82241--e82241, 2013.

\bibitem{zhu2016deep}
X.~Zhu, J.~Yao, and J.~Huang, ``Deep convolutional neural network for survival
  analysis with pathological images,'' in {\em 2016 IEEE International
  Conference on Bioinformatics and Biomedicine (BIBM)}, pp.~544--547, IEEE,
  2016.

\bibitem{chaddad2017predicting}
A.~Chaddad, C.~Desrosiers, M.~Toews, and B.~Abdulkarim, ``Predicting survival
  time of lung cancer patients using radiomic analysis,'' {\em Oncotarget},
  vol.~8, no.~61, p.~104393, 2017.

\bibitem{cui2020deep}
L.~Cui, H.~Li, W.~Hui, S.~Chen, L.~Yang, Y.~Kang, Q.~Bo, and J.~Feng, ``A deep
  learning-based framework for lung cancer survival analysis with biomarker
  interpretation,'' {\em BMC bioinformatics}, vol.~21, no.~1, pp.~1--14, 2020.

\bibitem{aerts2014decoding}
H.~J. Aerts, E.~R. Velazquez, R.~T. Leijenaar, C.~Parmar, P.~Grossmann,
  S.~Carvalho, J.~Bussink, R.~Monshouwer, B.~Haibe-Kains, D.~Rietveld, {\em
  et~al.}, ``Decoding tumour phenotype by noninvasive imaging using a
  quantitative radiomics approach,'' {\em Nature communications}, vol.~5,
  no.~1, pp.~1--9, 2014.

\bibitem{clark2013cancer}
K.~Clark, B.~Vendt, K.~Smith, J.~Freymann, J.~Kirby, P.~Koppel, S.~Moore,
  S.~Phillips, D.~Maffitt, M.~Pringle, {\em et~al.}, ``The cancer imaging
  archive (tcia): maintaining and operating a public information repository,''
  {\em Journal of digital imaging}, vol.~26, no.~6, pp.~1045--1057, 2013.

\bibitem{nakhleh2017diagnosis}
M.~K. Nakhleh, H.~Amal, R.~Jeries, Y.~Y. Broza, M.~Aboud, A.~Gharra, H.~Ivgi,
  S.~Khatib, S.~Badarneh, L.~Har-Shai, {\em et~al.}, ``Diagnosis and
  classification of 17 diseases from 1404 subjects via pattern analysis of
  exhaled molecules,'' {\em ACS nano}, vol.~11, no.~1, pp.~112--125, 2017.

\end{thebibliography}
}

\end{document}